\documentclass[useAMS,usenatbib]{mn2e}
\usepackage{natbib}
\usepackage{graphicx,amssymb,amsmath,epsfig}
\usepackage{epstopdf}
\usepackage{booktabs}
\usepackage{ulem,color}

\def\apjl{ApJL }

\def\apj{ApJ }

\def\aap{A\&A }
\def\nat{Nature }
\def\mnras{MNRAS }

\mathchardef\mhyphen="2D
 \definecolor{mypink1}{rgb}{0.9, 0., 0.6}

\title{Elliptical Accretion and Low Luminosity from High Accretion Rate Stellar Tidal Disruption Events}
\author[Svirski et al.]{ Gilad Svirski$^1$,  Tsvi Piran$^1$  and  Julian Krolik$^2$\\
$^1${Racah Institute of Physics, The Hebrew University of Jerusalem, Jerusalem 91904, Israel}
\\
$^2$ Physics and Astronomy Department, Johns Hopkins University, Baltimore, MD 21218, USA}
\begin{document}
\maketitle

\begin{abstract}
Models for tidal disruption events (TDEs) in which a supermassive black hole disrupts a star commonly assume that the highly eccentric streams of bound stellar debris promptly form a circular accretion disk at the pericenter scale.
However, the bolometric peak luminosity of most TDE candidates, $\sim10^{44}\,\rm{erg\,s^{-1}}$, implies that we observe only $\sim1\%$ of the energy expected from {radiatively efficient} accretion.
{Even the energy that must be lost to circularize the returning tidal flow  is larger than the observed energy. }
Recently, \cite{Piran+2015} suggested  that the observed optical TDE emission is powered by shocks at the apocenter between freshly infalling material and earlier arriving matter. This model explains the small radiated energy, the low temperature, and the large radius implied by the observations as well as the $t^{-5/3}$ light curve.  
However the question of the system's low {bolometric} efficiency remains unanswered. We suggest that the high orbital energy and low angular momentum of the flow 
make it possible for magnetic stresses to reduce the matter's already small angular momentum to the point at which it can 
{fall ballistically into the SMBH}  before circularization. 
As a result, the efficiency is only  $\sim1$--$10\%$ of a standard accretion disk's efficiency. Thus, the intrinsically high eccentricity of the tidal debris naturally explains why most TDE candidates are fainter than expected. 
\end{abstract}

\section{Introduction}\label{sec:intro}

The idea of circular accretion, offering a potentially extremely efficient radiation source, became ubiquitous with the successful application of the \cite{Shakura.Sunyaev.1973} model to active galactic nuclei and binary stellar systems. TDEs have also been traditionally envisioned as involving 
circular accretion.
However, recent TDE candidate observations very rarely reproduce these high efficiencies,
and recent detailed hydrodynamical studies of TDEs have indicated that most of their matter may stay on highly eccentric orbits for longer than previously thought.   These developments suggest the time has come to challenge circular accretion and examine alternative accretion pathways.

The classical description of a TDE was outlined by
\cite{Rees.1988} and \cite{Phinney.1989}. In this picture, immediately after the disruption half of the disrupted stellar debris remains bound and orbits the BH in highly eccentric streams.  On their first return to the pericenter region, their orbits are quickly circularized by shocks due to relativistic apsidal precession.   The mass fall-back rate $\propto{t^{-5/3}}$ translates into a similar lightcurve because the inflow time from the pericenter is short compared to the debris streams' orbital period. Based on this picture, a soft X-ray light curve comprising a bright peak followed by a $t^{-5/3}$ decay has long been taken to be the hallmark of a TDE, and a number of candidates have been found 
\citep[e.g.][]{Komossa+2004,Bloom+2011}.
Since then, even more TDE candidates have been found in the optical and UV
\citep[e.g.][]{Gezari+2009,Gezari.et.al.2012,Arcavi+2014,Holoien+2014,Holoien+2016,Holoien+2016oi}.

However, most properties of these TDE candidates are difficult to explain within the  classical picture. 
The observed  luminosity and temperature are much lower and the emission radius is much larger than expected.  Additionally, a major difficulty is that  the observed energy is significantly lower than the potential supply from accretion of a solar mass onto a SMBH {\citep[e.g.][]{Piran+2015}}.  In fact, for typical parameters, for which the pericenter $r_p$ of the tidally disrupted star is $\simeq10$--$30r_g$ ($r_g\equiv GM/c^2$), even  {the circularization energy, the } energy needed to be dissipated to form a circular accretion disk at $r_p$ is much larger than the observed radiated energy.

Possible resolutions to this problem discussed in the literature include photon trapping \citep{Begelman1979,Abramowicz1988,Krolik.Piran.2012}, a kinetic outflow carrying away orbital energy \citep{Ayal+2000,Ohsuga+2005,Sadowski+2014,Metzger.Stone.2015}, a low-mass outflow regulating the accretion rate \citep[e.g.][]{Poutanen+2007,Strubbe.Quataert.2009,Lodato.et.al.2009,Miller2015},
and radiation in an unobserved band such as the EUV.   However, whether by capturing the photons or limiting the accretion rate onto the BH (see also \citep{LoebUlmer1997}), most of these models  decouple the time-dependence of the luminosity from the fallback rate, and therefore lead to lightcurves different from the $t^{-5/3}$ power-law to which the data are often fitted. {Indeed, there are also observations indicating that a $t^{-5/3}$ decay often does {\it not} fit the data, both in the optical \citep{Brown+2016} and the X-ray band \citep{Miller+2015}.}

An additional indication of the severity of the circularization problem  has arisen from 
numerical hydrodynamics simulations of the mass return\footnote{{Note that while \cite{Bonnerot16} do show rapid circularization their initial conditions involve an elliptical orbit increaseing significantly the specific angular momentum of the flow as compared with the expected parabolic infall.}} .   Even within the first half an orbit of the most tightly-bound material, shocks near the orbits' apocenters begin to deflect the gas \citep{Rosswog.et.al.2009}.
Following the flow for $\simeq12$  orbits of the most tightly-bound material, \citealt{Shiokawa+2015} (hereafter S15) showed that, contrary to the expectation underlying the classical model, the returning matter does not quickly form an accretion disk around $r_p$.  
The gas does shock when it returns to the pericenter region, but, as predicted  \citep{Kochanek1994}, 
 this shock is weak; moreover, as a result of heating in the apocenter and nozzle shocks and other hydrodynamical effects, it lasts as a shock for only a few orbital periods of the most tightly-bound tidal debris.
Consequently, after passing through the near-pericenter region,  the matter continues to follow a highly-eccentric orbit and returns to an apocenter similar to the original orbit's.   Freshly infalling matter collides with this returning matter and forms additional shocks near the apocenter.  

\cite{Piran+2015} suggested  that  these outer  shocks, rather than  
accretion onto the SMBH, power the {optical} light output of TDEs.  In their model, the post-peak {optical} luminosity naturally decays $\propto t^{-5/3}$ because it is directly associated with mass fall-back, and the emerging radiation has the observed low luminosity, low temperature, and large radial scale. {Applied to the best observed TDE so far, ASSASN-14li \citep{Holoien+2016}, 
this model agrees nicely with all observed optical and UV features \citep{Krolik+16}.}

This model raises the question of the eventual fate of the 
matter. If it does eventually fall onto the SMBH, why does it not radiate a large amount of energy? 
Here, we propose a new accretion model,
in which a standard  accretion disk never forms.  
Instead, we analyze the dynamics of an intrinsically non-steady highly eccentric accretion flow.
Depending on the relative rates of energy and angular momentum loss, fluid on an elliptical trajectory may either 
circularize and form a standard accretion disk, or may enhance its eccentricity. In the latter case, upon losing sufficient angular momentum, fluid can cross the BH's effective potential barrier and plunge into the BH.
The orbital energy loss en route to the BH can then be considerably smaller than the one associated with standard disk accretion. 
We show that under  reasonable assumptions this scenario is likely, and it may explain why TDE candidates are fainter than expected. 

The feasibility of stable elliptical accretion disks with constant or rising eccentricity following a TDE was first discussed by \cite{Syer.Clarke.1992} (see also \cite{Lyubarsky.1994}), who suggested that elliptical disks may produce fainter TDEs due to longer accretion time-scales.  In contrast, we suggest here that TDEs are fainter because matter falls with \textit{less}, {in addition to} \textit{slower}, orbital energy loss.

\section{Orbital evolution in a highly eccentric disk}\label{sec:poc}

{Before considering the evolution of a highly eccentric disk, we turn to the question of whether such a disk forms at all, or whether the matter quickly circularizes instead.  The disrupted stellar matter moves initially on highly eccentric orbits.   Although the typical specific angular momentum  is of order $\sim \sqrt{ GM_{\rm BH} r_p}$, the specific energy of even the bound-most matter,  $\sim GM_{\rm BH} R_*/r_t^2$ ($r_t \geq r_p$ is the tidal radius), is  much larger than the {binding energy of a circular orbit with} this angular momentum. 
In order to form a circular disk at 
$r_p$, the orbiting gas must lose $\sim GM_{\rm BH}M_*/r_p$ of energy.   {As already noted, this is} more than a factor of ten times the total observed energy in most TDE.  }

{One could argue that this energy is somehow lost to heat, but not radiated away. The question is whether a suitable energy dissipation mechanism {exists}. The  {most commonly proposed mechanism} for circularization is shocks due to apsidal precession at the pericenter \citep{Rees.1988}.  However,  already in 1994   \cite{Kochanek1994} pointed out that these shocks at the pericenter are rather weak, and this prediction has been confirmed by 
recent numerical simulations. 
Unless $r_p\lesssim10r_g$ (corresponding, for $1M_\odot$ main sequence star events with pericenter near the tidal radius, to $M_{\rm BH} \gtrsim 1\times10^7M_\odot$), relativistic apsidal precession of the stellar debris during the disruption itself creates shocks near the stream apocenters rather than near $r_p$ \citep{Shiokawa+2015,Dai+2015}. {In addition,} the numerical simulations show that the shocks near the apocenter deflect the incoming matter in such a way that the motion near the pericenter cannot
be viewed simply as interacting streams coming from different directions,  Instead,  these simulations show that} {the dynamics of the tidal streams are subject to complex non-local hydrodynamic effects including multiple shocks as well as finite pressure gradients.   The end result is} an eccentric flow  whose
mean eccentricity is somewhat smaller than the initial mean eccentricity of the tidal debris (S15).   {Whilst these results have been obtained in a particular simulation, we believe that they are generic, and as such it is worthwhile to explore the implications of the resulting highly eccentric configuration that forms. } 

In the simulation reported in S15, the disk's mean eccentricity was $\simeq0.4$, but it may often be larger because the initial eccentricity of the tidal debris scales with the ratio of BH mass to star mass, $M_{\rm BH}/M_*$; this ratio was only 500 in their simulation, whereas it is expected to be $\sim10^6$ in typical TDEs.    To be precise, the
mass-weighted eccentricity of the tidal streams should be $\simeq [(M_{\rm BH}/M_*)^{1/3}-1]/[(M_{\rm BH}/M_*)^{1/3}+1] \simeq 1 - 2 (M_*/M_{\rm BH})^{1/3}$, which approaches unity as the mass ratio grows.    
The disk formed from the tidal debris therefore has approximately the same specific angular
momentum as a circular disk at $r_p$, but much higher energy. {We now consider the question of what happens to the matter in this highly eccentric disk after it has been established.}

\begin{figure}[h]
\includegraphics[width=\columnwidth]{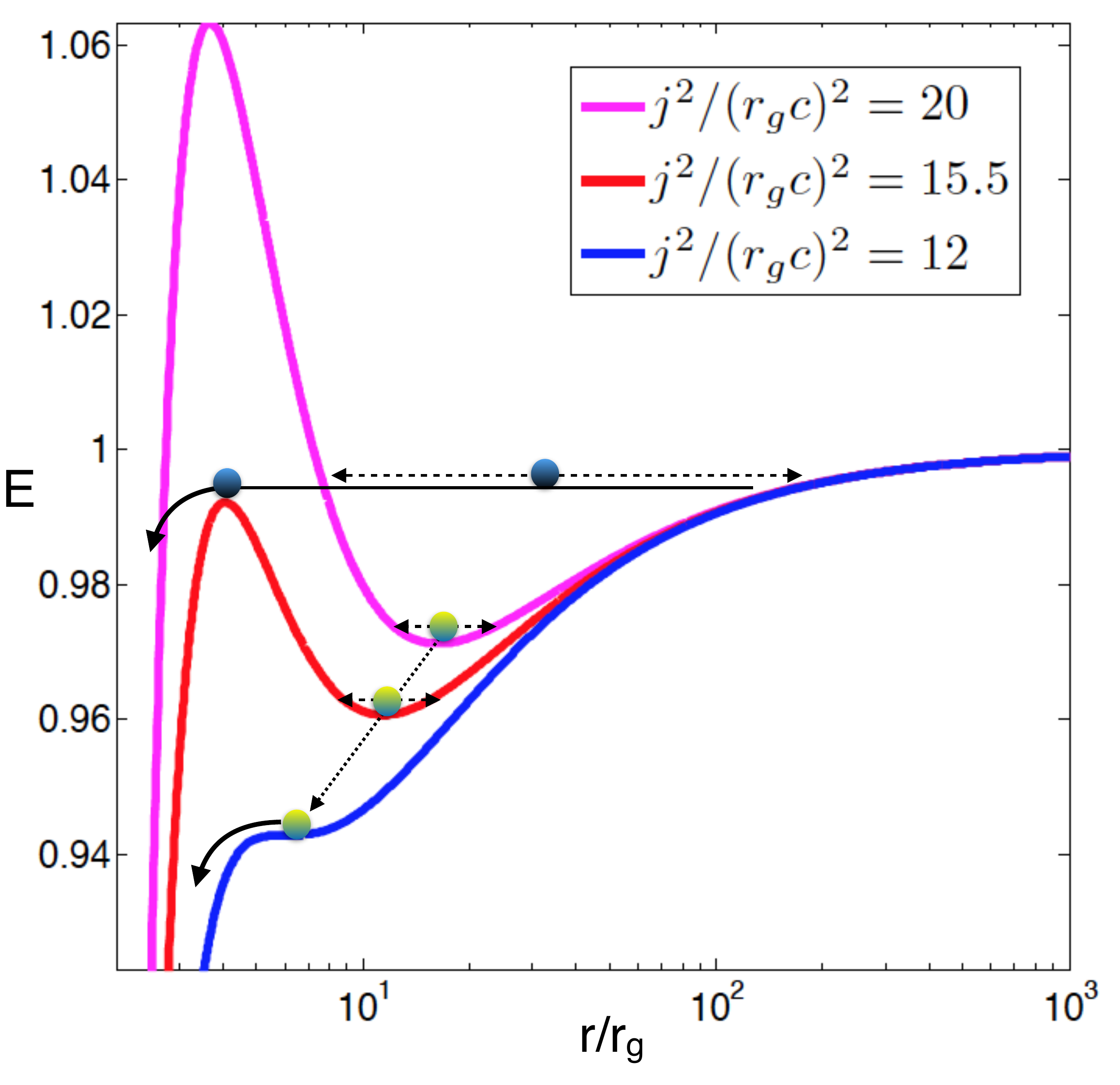}
\caption{
The specific energy $E$ associated with the Schwarzschild effective potential 
for three values of specific angular momentum.  Dashed lines with arrows indicate bounded orbital motions, solid lines 
indicate direct infall onto the black hole.
The fluid element represented by a blue ball follows an elliptical trajectory (black dashed line); upon losing a small amount of angular momentum and almost no energy, it plunges into the black hole (black solid line).
The fluid elements represented by green balls are on circular orbits  and can fall into the BH only after losing a significant amount of energy.   The dotted line depicts their motion
 from one circular orbit to the other.  In this ``standard" circular accretion, plunge into the black hole is possible only when
the angular momentum has been reduced to that of the ISCO.
}
\label{Fig:accretion}
\end{figure}

{Highly eccentric accretion disks around black holes differ qualitatively from circular disks.  In a circular disk, if}
{the debris are to reach the innermost stable circular orbit (ISCO), {they must do so by gradually losing both energy and angular momentum through the action of magnetic stresses.}}  {However, in an eccentric disk,  particles can pass through the event horizon {\it with no energy loss} if their trajectories have sufficiently small angular momentum} (see Fig \ref{Fig:accretion}). In the context of an eccentric accretion disk, it is therefore of considerable interest to examine carefully whether gas streams reach such low angular momentum before or after their binding energies become comparable to the ISCO binding energy.

For a particle on a parabolic trajectory, i.e., with specific energy $E=1$, to fall all the way to the ISCO radius of a Schwarzschild black hole its specific angular momentum $j$ must be $<4r_gc$ (see Fig \ref{Fig:accretion}).
By comparison, a parabolic orbit with pericenter $r_p$ has a specific angular momentum
\begin{equation}\label{eq:l}
j_p=\frac{r_p}{\sqrt{r_p/(2r_g)-1}}.
\end{equation}
Substituting a fiducial $r_p=15r_g$, as expected in a TDE of $M_{\rm{BH}}\approx3\times10^6M_{\odot}$ (e.g. Piran et al. 2015), yields $j_p=5.9r_gc$. 
Thus, such a fluid element must lose only $30\%$ of its angular momentum in order to cross the potential barrier of a nearly parabolic orbit.
We now examine the possibility that the loss of $30\%$ of the angular momentum is accompanied by an orbital energy loss much smaller than the $6\%$ of rest-mass energy associated with circular accretion onto a Schwarzschild BH.
In other words, we ask whether the orbits may evolve in such a way that they are able to plunge directly into a black hole even while the associated semi-major axes remain much larger than $r_{\rm ISCO}$.
Because orbital energy dissipation associated with apsidal precession may become important only when a fluid element passes within $\simeq 10r_g$ of the black hole,
we initially ignore its effects; we return to this topic after exploring how fluid angular momentum evolves under the influence
of internal MHD stresses alone.

\subsection{Energy and angular momentum transport by MHD turbulence}

As the tidal debris travels along its highly elliptical orbits, one might reasonably expect the now-familiar exponential growth
of the magneto-rotational instability \citep{BalbusHawley.1991} 
Because the fastest-growing modes of the instability grow at a rate $\sim\Omega$, where $\Omega$ is the {\it local}
orbital shear \citep{Balbus.Hawley.1998}, this growth might be rather different when both the orbital frequency and the
shear vary drastically around the orbit, as they do in an eccentric disk, rather than remaining constant as in a conventional 
circular disk.   We set aside this question to be answered later by numerical simulations.    

The rate at which a fluid element's angular momentum changes is given by the divergence of the angular momentum flux through that fluid element. In steady-state accretion disks, it is simplest to analyse the flow from an Eulerian point of view, in which angular
momentum and energy  flow radially through the disk.
Thus, any divergence in flux is due entirely to radial gradients, and these apply to fluxes associated with both magnetic stresses and mass flow.   In the case of eccentric  
accretion flows, it is more convenient to think instead in terms of a Lagrangian picture
which is more appropriate to following the orbital evolution of an individual fluid element.   

In a circular accretion disk, the dominant off-diagonal element of the Maxwell stress tensor 
is the $\hat{r}$-$\hat{\phi}$
element, associated with conveying angular momentum in the radial direction.   This is the only element with a non-zero long-term mean value; 
both of the other two elements fluctuate symmetrically in sign.   By contrast, in an eccentric disk the
orbital velocity has both radial and azimuthal components and varies in a direction $\hat{n}$ perpendicular  to the local orbital
tangent $\hat{t}$.    We will therefore assume that the local Maxwell stress tensor follows suit: its only interesting off-diagonal
element is the $\hat{n}$-$\hat{t}$ element.    To evaluate its magnitude, we will make two further assumptions: that
$\alpha_{\rm mag}\equiv\langle-2B_{\hat{n}\hat{t}}\rangle/\langle B^2\rangle$ 
is the usual $\simeq0.4$ \citep{Hawley+2011};
and that the scale of the gradient in the $\hat{n}$ direction is $\sim r$.

With these assumptions, the rate of specific angular momentum loss  is
\begin{equation}
{\partial j\over\partial t}\simeq
{1\over r\Sigma(r)} \int\,dz\,r |\hat{r}\times\hat{t} | \alpha_{\rm mag} {B^2 \over 4\pi}.
\end{equation}
The factor $|\hat{r}\times\hat{t}|=(1+e\cos\phi)/(1+e^2+2e\cos\phi)^{1/2}$ is necessary to convert momentum in the $\hat{t}$ direction into angular momentum.
For most angles $\phi$, $|\hat{r}\times\hat{t}|\sim O(1)$.   However, for directions toward the apocenter it is generally $\sim(1-e)^{1/2}\sim(M_*/M_{_{\rm BH}})^{1/6}$ because the orbital tangent at large distances is almost radial except for a very short stretch
at the apocenter itself.

The rate at which internal stresses turn orbital energy into heat is the rate of dissipation associated with the turbulence.
In a time-steady circular disk, this rate is the rate of shear times the stress across the shear, i.e., $(\partial\Omega/\partial\ln r)\int\,dz\,T_{r\phi}$.   In a non-steady elliptical 
flow, this relation is unlikely to apply exactly, but we will adopt it as an order of magnitude estimate, i.e., the heating rate per unit mass is
\begin{equation}
{\partial{E}\over\partial{t}}\simeq{\Omega(r)\over r\Sigma(r)}\int\, dz\,r\alpha_{\rm{mag}}{B^2\over4\pi},
\end{equation}
with the important distinction that $\Omega(r)\equiv\dot\phi=j/r^2$ is the {\it local} orbital frequency , {\it not} $2\pi\times$ the inverse of the orbital period.
Note that it does not matter whether the work takes the form of a torque changing rotational kinetic energy or a force
changing translational kinetic energy, so the geometric factor $\hat{t}\times\hat{n}$ is unnecessary. Note, too, that
dissipation of orbital energy into heat is not in itself a loss of energy; energy is removed from the flow only when it actually leaves,
whether in the form of escaping photons or escaping matter.

{Consider now a fluid element that is on an elliptical trajectory with a pericenter $r_p$,  the pericenter of the stellar orbit.} If we freeze the torque at the value pertaining to a particular location $(r,\phi)$ in the orbit, the time required to reduce the
initial angular momentum to its critical value, $j_{\rm{crit}}$, is
\begin{eqnarray}
t_{_J}(r)&=&\frac{\Delta j}{\partial{j}/\partial{t}}=\frac{\left[(j_p-j_{\rm crit})/c^2\right](v_{\rm circ}^2/v_A^2)(r/r_g)}{\alpha_{\rm{mag}}|\hat{r}\times\hat{t}|},
\end{eqnarray}
for $v_{\rm{circ}}$ the circular orbital speed at radius $r$ and $v_A$ the local Alfven speed.    The factor $1/v_A^2$ should
be interpreted as
${\int\,dz\,\rho(z)/\int\,dz\,B^2/(4\pi)}$.
 $j_{\rm crit}=4r_gc$  for non-spinning BHs; it is somewhat less for larger spin parameters.   Writing $j_{\rm{crit}}={\cal{J}}_{\rm{crit}}r_gc$ and approximating $j_p\simeq(2r_p/r_g)^{1/2}r_gc$, the preceding expression can be rewritten as
\begin{eqnarray}
t_{_J}(r)=\frac{\sqrt{2} [1-{\cal J}_{\rm crit}/(2r_p/r_g)^{1/2}](r_p/r_g)^{1/2}(r/r_g)r_g/c}{\alpha_{\rm mag}(v_A^2/v_{\rm circ}^2) |\hat{r}\times\hat{t}|} \ ,
\end{eqnarray}
The time associated with reducing the initial specific orbital energy to $-GM_{_{\rm BH}}/(2r_p)$, the energy of a circular orbit at radius $r_p$ is
\begin{equation}
t_{_E}(r)=\frac{\Delta E}{\partial E/\partial{t}}=\frac{(r_g/r_p)^{3/2}(r/r_g)^3r_g/c}{2^{3/2}\alpha_{\rm mag}(v_A^2/v_{\rm circ}^2)}.
\end{equation}

S15 found that the disk aspect ratio $H/R$ varies slowly as a function of radius because adiabatic compression of fluid elements as they move inward raises their temperature. The same compression also increases the flow's optical depth and lengthens its cooling time, enforcing adiabatic behavior.    When $H/R$ is nearly constant, and the usual relations between magnetic intensity and
internal pressure hold, $v_{\rm{circ}}/v_A$ should likewise change only weakly with radius.
Thus, both characteristic timescales are shortest near pericenter.

At any particular radius, the ratio between these two times is
\begin{equation}
t_{_J}/t_{_E}\sim2[1-{\cal J}_{\rm{crit}}/(2r_p/r_g)^{1/2}]|\hat r\times\hat{t}|^{-1}(r_p/r)^2 \ .
\end{equation}
The factor within square brackets is always order unity; the geometric factor is $\sim O(1)$ near pericenter, but grows to $\sim O(1-e)^{-1/2} \sim (M_*/M_{_{\rm BH}})^{-1/6}$  
toward apocenter.   However, the final factor falls to $\sim4(1-e)^{2}\sim (M_*/M_{_{\rm{BH}}})^{2/3}$ near apocenter.  Consequently, the overall product is comparable to unity near pericenter, but decreases drastically at larger distance from the BH.  Thus, throughout the orbit, progress toward reaching the critical angular momentum is at least as fast as the effective rate of progress toward the circular-orbit binding energy and can be considerably faster.

To estimate how many orbits are required to reach either goal consider the product of $t_{_J}(r)$ or $t_{_E}(r)$ with $\Omega(r)$:
\begin{equation}
\frac{1}{j_p-j_{\rm{crit}}}{dj\over d\phi}=\left(t_{_J}\Omega\right)^{-1}
\end{equation}
and
\begin{equation}
\frac{2r_p}{r_g{c^2}}{dE\over d\phi}=\left(t_{_E}\Omega\right)^{-1}.
\end{equation}
The ratio between the angular momentum removed while traversing a radian of azimuthal angle and 
 the angular momentum that must be lost in order to plunge to the BH is:
\begin{eqnarray}
\left(t_{_J}\Omega\right)^{-1}=\frac{
\alpha_{\rm{mag}}(v_A^2/v_{\rm{circ}}^2)(r_p/r_g)^{1/2} |\hat{r}\times\hat{t}|(r/r_p)}{\sqrt{2}[1-{\cal{J}}_{\rm{crit}}/(2r_p/r_g)^{1/2}](j/r_g c)}.
\end{eqnarray}
Over the
same radian of azimuthal angle, the binding energy is increased by
\begin{equation}
\left(t_{_E}\Omega\right)^{-1}= \left(t_{_E}\Omega\right)^{-1}= 2 (r_p/r) \alpha_{\rm mag} (v_A^2/v_{\rm circ}^2)
\end{equation}
in units of $GM_{\rm BH}/(2r_p)$.

These two expressions may be approximately integrated over the entire orbit using 
\begin{equation}
r(\phi)=\frac{(1+e)r_p}{1+e\cos\phi}
\end{equation}
and dividing the full range of azimuthal angle into three segments.   Here $e$ is the eccentricity of the fluid orbit.  For the roughly half of azimuthal angle near pericenter, $r/r_p\simeq1$ to within a factor of 2; for angles within a radian of apocenter but not extremely close to the apocenter, $r/r_p\simeq2(1+e)/(e\phi^2)$; for angles so close to apocenter that $|\phi-\pi|<\sqrt{2(1-e)/e}$, $r/r_p\simeq(1+e)/(1-e)$.     
The total angular momentum loss over an orbit is dominated by the region very close to the apocenter, and its ratio to the amount necessary to fall into the BH is
\begin{equation}\label{eqn:jfrac}
\frac{\delta j}{j_p-j_{\rm crit}}\simeq\frac{{\sqrt{2}}\alpha_{\rm mag}(v_A^2/v_{\rm circ}^2)(r_p/r_g)^{1/2} } {[1-{\cal J}_{\rm crit}/(2r_p/r_g)^{1/2}](j_p/r_g c)(1-e)^{1/2}}.
\end{equation}
The analogous fractional energy loss per orbit is dominated by radii near the fluid pericenter:
\begin{equation}\label{eqn:efrac}
\frac{\delta E}{GM_{_{\rm{BH}}}/(2r_p)}\simeq\sqrt{2}\pi(j_p / r_g c)^{-1}\alpha_{\rm{mag}}(v_A^2/v_{\rm circ}^2)(r_p/r_g)^{1/2}.
\end{equation}

MHD simulations of fully-developed MRI-driven turbulence 
 find $\alpha_{\rm{mag}}\simeq0.4$, while $v_A^2/v_{\rm{circ}}^2\simeq0.01$--0.1.    The factor $(r_p/r_g)^{1/2}$ is typically several.   Combining these factors with $(1-e)^{-1/2}\sim (M_{_{\rm{BH}}}/M_*)^{1/6}\sim10$, we expect that {$\sim 10$} orbits may be necessary to remove so much angular momentum that a stream can fall directly into the BH.  
 The expression for the energy loss is very similar to that for angular momentum loss except for a factor of $(1-e)^{-1/2}\sim10$ (the other remaining factor is $\pi [1-{\cal{J}}_{\rm{crit}}/(2r_p/r_g)^{1/2}] \sim 1$).  Thus, the rate of progress toward losing enough angular momentum to be accreted is likely to be $\sim 10\times$ faster than the rate at which the orbits evolve toward a semi-major axis $\sim r_p$; as a result, at the time when streams plunge through the ISCO, their orbital semi-major axes are $\gg r_p$, which itself is considerably greater than $r_{\rm ISCO}$.

It therefore appears that much of the matter may reach the BH at a time when its orbital energy has been reduced by only a fraction of the binding energy at $r_p$, which is in turn,  for our fiducial $j_p$, three times smaller than the binding energy of an orbit
near the ISCO.   The efficiency is then only $\sim3\%$ of the classical one. This efficiency also holds for non-fiducial cases
because the ratio of the dissipated energy to the classical efficiency is
$\sim \pi [1-{\cal{J}}_{\rm crit}/(2r_p/r_g)^{1/2}](1-e)^{1/2} (r_{\rm ISCO}/r_p)$.   Taken together, the last two factors scale
$\propto M_{\rm BH}^{1/2}$, but the first factor decreases from unity for small $M_{\rm BH}$ to zero when $M_{\rm BH} \gtrsim
3 \times 10^7 M_{\odot}$.   Thus, the overall product varies only slowly for masses up to $\sim 3 \times 10^7 M_{\odot}$,
where it becomes even smaller.
The fraction of rest-mass energy going into heat may therefore be just a few percent of typical steady-state radiative efficiencies.

In fact, only a part of this heat may be radiated, and the photons are most likely to emerge from a cooler
part of the disk than the place in which the gas is heated.  Near the pericenter, where the material is hottest,
the surface temperature is likely to be high enough for the output to be primarily in the soft X-ray band.  However,
the cooling time there is so long compared to the time spent there that little energy is radiated near pericenter.
This is because the laminar orbital flow in a highly eccentric disk with constant $H/R$ causes the density to scale
$\propto r^{-3}$ so that $t_{\rm cool} \Omega\propto r^{-3}$, while, as shown by \citet{Piran+2015}, the cooling
time is comparable to the orbital frequency at the apocenter when the orbit extends out to $r \sim a_{\rm min}$.
Moreover, adiabatic expansion reduces the internal energy available to cool at larger radii where photon diffusion is more rapid.   Consequently, the fraction of the heat created at $r\sim r_p$ that is radiated each orbit is only
$\sim (1-e)^2$, so that the actual {\it radiated} efficiency may be only $\sim O(0.1)\times$ the heating efficiency.
Although concatenating these several uncertain factors creates a quantity with even larger uncertainty, we
suggest that the radiative efficiency associated with dissipation of MHD turbulence may be only $\sim0.01$--$0.1\times$
the generic radiative efficiency of black hole accretion, i.e. roughly comparable to or slightly greater than the
radiative efficiency of the apocenter shocks. The net result is that when the eccentric orbits stretch out all the way to $r \sim a_{\rm min}$, the heat generated at small radii by dissipation
of MHD turbulence is effectively mixed with the heat generated at these larger radii by the outer shocks.  The two
heat sources are then indistinguishable observationally, both contributing to optical/UV light.

\subsection{Validity of assumptions: the saturation amplitude of MHD turbulence, the effects of apsidal precession, and outward angular momentum transport}

To close this section, we return to two assumptions that might be questioned {and to a global issue, closely related to one of these assumptions, that has been ignored in the essentially local treatment  of orbital evolution we have employed.}

The first {assumption made } is that $v_{\rm circ}^2/v_A^2$ remains approximately constant
around the orbit, i.e. that $v_A$ falls with radius $\propto r^{-1}$.   If it were to fall with radius at least as rapidly as $\propto r^{-3/2}$, angular momentum loss would be dominated by the pericenter rather than the apocenter region; energy loss would remain dominated by the pericenter.    In this case, the efficiency of accretion from a highly eccentric stream is limited to $\le GM_{\rm{BH}}M_*/r_p$, which is always less than the classical radiative efficiency of time-steady nearly-circular flow through the ISCO.    Alternatively, if $v_A$ were to fall with radius more slowly than $\propto r^{-1}$, angular momentum loss would become even more dominated by the apocenter region, but provided $v_A \propto r^n$ with $n<1$, energy loss would continue to be primarily near the pericenter.    In this case, the rate of angular momentum loss would remain several times greater than the rate of energy loss.

In addition, and perhaps more importantly, we have up to this point ignored relativistic apsidal precession.   Although reasonable at large radius, this assumption may fail once the fluid's pericenter has been reduced from $r_p$ to $\lesssim 10r_g$, where test-particle orbits have significant precession angles and self-intersect at $\lesssim 100r_g$ \citep{Dai+2015}.  Substantial apsidal precession can have several consequences.

The first follows directly from the shift in the position of a stream's orbital pericenter.   Apsidal precession rotates the azimuthal angle of the pericenter in the direction of the stream's orbit; conversely, the stream's trajectory at the original pericenter azimuth is at a radius larger than $r_p$.   Because the precession angle is $\propto r^{-1}$, fluid elements traversing orbits with smaller angular momentum therefore press outward against orbits with slightly greater angular momentum near the inner fluid's former pericenter.  The ensuing radial pressure gradient does negative work on the inner stream, decreasing its semi-major axis.

The second is that if the region inside the nested elliptical orbits is relatively evacuated, as it might be if lower angular momentum orbits fall rapidly into the black hole, a large precession angle can cause a stream to cross through the central region and shock against the orbiting fluid on the opposite side.   When two streams strike each other supersonically, a pair of shocks forms, one acting on each stream.   The region between the two shocks is filled with a pair of post-shock regions separated by a contact discontinuity. Post-shock, the fluid from both streams rises in pressure as shock dissipation transforms orbital energy into heat, while momentum exchange in the shock pair takes angular momentum from one shocked stream and gives it to the other.  In this case, because there is the large inertia of the disk body behind the outer, larger angular momentum material, the sense of the momentum exchange will be largely to remove momentum from the inner stream, deflecting it inward, toward the evacuated cavity, and diminishing its angular momentum.    Because the fluid pressure high above the disk midplane is also low, some of the inner stream may be deflected upward.   The net effect of upward deflection is to place the stream on an orbit highly inclined relative to the orbital plane, but also with diminished angular momentum.
When matter on this sort of orbit next passes through the orbital plane, it may be directly accreted onto the black hole.   Such effects have already been seen in some numerical simulations \citep{Sadowski+15}.
Thus, shocks driven by apsidal precession can transfer angular momentum from low specific angular momentum material to matter with higher specific angular momentum, performing a role analogous to the MHD stresses responsible for angular momentum transport in conventional circular disks.

The orbital eccentricity depends on both energy and angular momentum: $e^2 = 1 + {\cal E} j^2/(GM)^2$, for specific energy ${\cal E}$ and specific angular momentum $j$.    We have just argued that shocks generally reduce $j$ for the innermost streams, but they also make ${\cal E}$ more negative by dissipating orbital kinetic energy into heat.   It is possible that in some cases, $e$ actually decreases if the energy change outweighs the angular momentum,
whereas in other cases $e$ may change in the opposite direction.   Even when the gas moves to a more circular orbit of small radius, little radiation may result.
Although it has been heated strongly, escape of any radiation to infinity is slow because the optical depth is very large.  We earlier argued that heat created by MHD dissipation near the fluid's orbital pericenter was more likely to be radiated when the fluid element has travelled to apocenter than near the place where it was generated.   Fluid elements on nearly-circular orbits of small radius never go far enough outward to radiate efficiently.

The last consequence of apsidal precession-driven shocks has to do with the matter from the outer material that is also shocked.   Its eccentricity generally decreases because it both gains angular momentum and loses orbital energy.   The most immediate effect of this partial circularization is that the increase in $j$ in this matter increases the orbital shear between it and matter farther out.   One might then expect a growth in the amplitude of the MHD turbulence and a greater internal stress.    The end-result would be outward angular momentum transport by MHD stresses much like that in ordinary circular accretion disks.  Ultimately, the angular momentum of accreted mass is transferred all the way to the outer edge of the disk.   Once there, if there is no companion, as is likely for black holes in galactic nuclei, it must reside in this outermost material.   If that matter's orbit had been eccentric, it grows more circular; if it is already circular, it moves to larger radius.

The net effect of this angular momentum transport is surprisingly modest, even if only a small fraction of the initial bound mass ends up holding the angular momentum initially possessed by all the rest.    The reason is partly that the total angular momentum content of a tidal debris disk is relatively small, precisely because the orbits are so eccentric, and partly because the specific angular momentum for a circular orbit at radius $r$ is $\propto r^{1/2}$, so that angular momentum transported a large distance outward creates a relatively small change in those distant orbits.

To support a more quantitative argument, we remind the reader that the initial state of the disk is one in which most of the mass has orbits with semi-major axes $\gtrsim a_{\rm min}$ and typical eccentricity $\simeq 1 - 2 (M_*/M_{\rm BH})^{1/3}$.  When angular momentum is removed from some of the matter, allowing it to plunge directly through the ISCO and into the black hole, that angular momentum is transported outward through MHD torques and delivered to the remaining matter.  The natural scale for estimating its effect is the specific angular momentum required to support a circular orbit at $r=a_{\rm min}$, $j_{\rm circ}$.   The ratio between the initial specific angular momentum and the circular orbit value is
\begin{equation}
j_0/j_{\rm circ} = (1 - e^2)^{1/2} \simeq 2^{3/2} (M_*/M_{\rm BH})^{1/3},
\end{equation}
assuming $1-e \ll 1$.   If a fraction $f_M$ of the gas mass is accreted onto the black hole through direct plunge orbits, it must transfer a fraction $f_j$ of its initial angular momentum to the remaining gas.    As a result, the specific angular momentum of the mass remaining in the disk becomes
\begin{equation}
j = j_0 (1 + f_j f_M)/(1 - f_M).
\end{equation}
For our fiducial conditions (a $1 M_{\odot}$ main sequence star disrupted by a $10^{6.5}M_{\odot}$ Schwarzschild black hole), $j_0/j_{\rm circ} \simeq 0.019$ and $f_j \simeq 0.32$.   Even if 90\% of the original mass is accreted by plunging, the remaining matter still has $\simeq 1/4$ the angular momentum necessary to support a circular orbit at $a_{\rm min}$, and therefore an eccentricity $\simeq 0.87$.    Thus, the remnant disk remains quite eccentric until it has been reduced to a very small fraction of its initial mass.

{On the other hand, the mean specific angular momentum of the remaining material does become significantly larger than $j_0$ once $\sim 1/2$ of the disk mass has been accreted.  For accretion with little energy loss to continue, angular momentum transport, whether by MHD stresses or hydrodynamics, must keep $j$ of lower energy material far enough below the mean that the time for $j$ to fall below ${\cal J}_{\rm crit}$ is shorter than the orbital energy loss time.  {Because MHD stresses in conventional circular disks readily create a specific angular momentum profile in which $j$ rises outward, and because angular momentum exchange in shocks also leads to such a condition, this is a plausible, though hardly proved, state for evolving eccentric disks.} At the point when this is no longer the case,} the accretion mode for the remaining gas changes to the conventional one in which most of the initial orbital energy is lost through dissipation and radiation.  {However, even in this stage,} the luminosity remains very low, $\sim \alpha_{\rm mag} (v_A^2/v_{\rm circ}^2) (1 - f_M) \eta (M_* c^2/2)/t_0$, {because the characteristic timescale for energy loss is the orbital period, which is much longer than $t_0$ until dissipation has removed most of the orbital energy.} Here $\eta$ is the usual radiative efficiency for circular disks around black holes.  Thus, the resulting luminosity is reduced relative to the conventional expectation by the product of one factor ($v_A^2/v_{\rm circ}^2$) expected to be $\lesssim 10^{-2}$ in this context and another ($1-f_M$) {is always well less than unity.}

As these complicated and partially conflicting considerations indicate, the net outcome of low angular momentum gas undergoing relativistic apsidal precession is difficult to gauge without a full simulation of the relevant relativistic hydrodynamics.   Such an effort is well beyond the scope of the present paper, whose purpose is to present a first exploration of these ideas.

\section{Summary}\label{sec:conc}
 
Except for those events in which the stellar pericenter lies within $\simeq 10r_g$, shocks near the initial stream apocenters create an irregular, lumpy, and highly eccentric accretion flow in which individual streams can no longer be distinguished \citep{Shiokawa+2015}.
We have shown { here } that reasonable estimates of the rates of angular momentum loss and energy loss due to MHD effects lead to the conclusion that angular momentum is lost sufficiently rapidly to 
bring streams near or below the critical angular momentum at which they plunge directly into the SMBH, losing only a small fraction of their orbital energy.   The actual energy dissipated is equivalent to $\sim3\%$ of the expected, $GM_{\rm{BH}}M_*/r_{\rm{ISCO}}$, and only a fraction of this heat is likely to be radiated before the matter plunges into the black hole.   
Thus, highly eccentric accretion flows may radiate little enough to explain the comparatively small radiation fluence observed from most TDE candidates.

{We have presented estimates indicating that a sizable fraction of the bound mass following a tidal disruption can be accreted in this weakly-radiative fashion, but a number of complex hydrodynamical issues must be considered in greater detail for these conclusions to be made firmer.}
Apsidal precession  may cause additional dissipation in shocks once the gas's pericenter is
$\lesssim 10r_g$.  Depending on the strength of the shocks and on the ratio of energy loss to angular momentum loss in individual fluid elements, both approximate circularization and rapid plunging into the black hole may result.   {Both MHD stresses and shocks may transport angular momentum and energy globally through the flow, altering orbital shapes.} Reducing the quantitative uncertainty involved in these estimates will therefore require considerably more detailed analytical and numerical calculations.
Although these estimates were motivated by TDEs, this result  may also be applicable to other systems whenever highly eccentric flows arise, e.g. in a disk form from the fall back of tidal debris in a compact binary merger.

Before concluding, we recall that the observed optical/UV emission {can be} powered by the kinetic energy of the returning flow {as it shocks against} material orbiting at the apocenter of the most bound material \citep{Piran+2015}.   This emission is expected to peak $\sim5$ characteristic orbital periods after the disruption itself; after $\sim1$--2~orbital period at peak luminosity, it should diminish roughly $\propto t^{-5/3}$ \citep{Shiokawa+2015,Piran+2015}.   
Because {cooling is slow at radii inside} the apocenter, the heat dissipated by MHD turbulence near $r_p$ becomes spectrally mingled with the heat generated in the apocenter shocks.      Depending on how rapidly the MHD turbulence saturates, {its dissipation rate} may peak at about the same time {as the apocenter shocks'}, but may {last} as much as an order of magnitude longer.   Because its total energy is comparable, its contribution to the peak luminosity {would then be only $\sim0.1\times$ as large}, but it may also contribute to a longer timescale tail of the lightcurve. 
{Although the luminosity is well below the level expected on the basis of efficient accretion, the combination of apocenter shocks and inefficient accretion can still result in near-Eddington heating rates \citep{Piran+2015} .   When that is the case, the apocenter region, where most of the radiation escapes, may exhibit some of the mechanisms previously discussed in other contexts in which material near supermassive black holes is subjected to near-Eddington radiation fluxes \citep{ProgaKallman04,Strubbe.Quataert.2009,Lodato.et.al.2009,LaorDavis2014,Miller2015,Roth2016}. 
Future, more detailed, work on dissipation in highly elliptical accretion flows will support more specific predictions of the kinds of radiation such flows emit.}

This work was supported by the grants:   Advanced ERC TReX, I-CORE  1829/12,  ISA  3-10417  (TP),  NSF  AST-1028111 and NASA/ATP  NNX14AB43G (JHK).

\bibliographystyle{mnras}

\end{document}